'Some Thoughts on the Fluid Equation of Cosmology'

J. Dunning-Davies,
Department of Physics,
University of Hull,
England.

j.dunning-davies@hull.ac.uk

Abstract.
   The well-known fluid equation of cosmology is examined with a view to elucidating the precise conditions under which it is applicable.

From the Second Law, it may be deduced that both an absolute temperature, $T$, and an entropy, $S$, exist and
$$TdS = d'Q,$$
where $d'Q$ represents an increment of heat. This derivation and final equation strictly refer only to processes which may be run in reverse. When other processes are taken into account, it is generally accepted that the equality is replaced by the inequality
$$TdS \geq d'Q,$$
although questions relating to the existence of an entropy in such situations remain. An *adiabatic* process is one during which there is no heat flow. Hence, for such a process,
$$d'Q = 0.$$
If such a process may be run in reverse, $dS = 0$. However, if that is not the case, present thinking indicates that $dS \geq 0$. These two expressions follow since absolute temperature is an essentially positive quantity.

The above relations give the basic thermodynamic framework derived from the basic laws of the subject and within which other branches of physics must operate if they are to obey the laws of thermodynamics.

It seems that the basis of cosmology is firmly rooted in the above thermodynamics. Although there appear to be several ways of approaching the topic, they all seem to amount to the same thing in the end. According to books such as Liddle[1] and Ryden[2], the starting point is the Friedmann equation
$$\frac{\dot{a}^2}{a^2} + \frac{kc^2}{a^2} = \frac{8\pi G}{3}\rho,$$
where Liddle's notation has been adopted.

It is then pointed out that this equation is of little use unless coupled with another equation describing the evolution in time of the density, $\rho$, of material in the Universe. This is achieved via thermodynamic reasoning. The starting point is quite specifically the First Law of Thermodynamic:
$$d'Q = dU + pdV.$$
This is applied to an expanding volume $V$ of unit comoving radius but, therefore, with physical radius $a$. Then, $dU$ represents an increment of internal energy, where
$$U = \frac{4\pi}{3}a^3\rho c^2,$$
so that
$$dU = 4\pi a^2 \rho c^2 \frac{da}{dt} + \frac{4\pi}{3}a^3 \frac{d\rho}{dt}c^2.$$
Also
$$dV = 4\pi a^2 \frac{da}{dt}.$$
Hence, substituting into the above equation for the First Law gives
$$d'Q = \frac{4\pi}{3}a^3 c^2 \left( \frac{d\rho}{dt} + \frac{3\rho}{a}\frac{da}{dt} + 3\frac{p}{ac^2}\frac{da}{dt} \right)$$
$$= \frac{4\pi}{3}a^3 c^2 \left( \dot{\rho} + 3\frac{\dot{a}}{a}[\rho + p/c^2] \right).$$

If, and this is a crucially important point, the process under discussion is an adiabatic process, then $d'Q = 0$ and
$$\dot{\rho} + 3\frac{\dot{a}}{a}\left(\rho + p/c^2\right) = 0.$$
Hence, this so-called fluid equation holds only for adiabatic processes!

It is important to realise that this derivation does not depend on properties of the entropy. The derivation of the fluid equation depends only on the First Law of Thermodynamics. It is the Second Law which introduces the thermodynamic concept of entropy. The equation deduced is the familiar
$$TdS = d'Q,$$
and this is deduced by considering a thermodynamic cycle in which the end point and starting point are identical thermodynamically. Hence, the above fluid equation will hold automatically for an isentropic ($dS = 0$) process also. Further, it is felt[3] that, if an adiabatic process which may not be run in reverse is considered, this result for an isentropic process will be modified to
$$dS \geq 0.$$
The extension of this notion to all thermodynamic processes is an assumption and the generalisation of the basic equation representing the Second Law to
$$TdS \geq d'Q$$
is yet another basic assumption, as has been pointed out by Meixner[4]. Again, as Meixner has made quite clear, not only is this another basic assumption but, underlying it, is yet another fundamental assumption relating to the existence of entropy in situations which are not reversible.

Originally, the purpose of this short note was to draw attention to the conditions under which the fluid equation holds. However, it is obvious that other quite fundamental issues relating to the basis of thermodynamics have appeared and initial indications are that a return to examining the foundations as put forward by the founding fathers of the subject- people such as Carnot, Thomson, Tait and Clausius - could prove beneficial. The results of such an investigation will provide the material for future contributions.

**References.**

1. A. Liddle, 1999, *An Introduction to Modern Cosmology,*
   Wiley, Chichester.

2. B. Ryden, 2003, *Introduction to Cosmology,*
   Addison-Wesley, New York.

3. J. Dunning-Davies, 2007, *Concise Thermodynamics*, 2$^{nd}$ edition
   Horwood, Chichester.

4. J. Meixner, 1969, Arch. Rat. Mech. Anal. **33**, 33.